# *Evidence for a Kondo destroying quantum critical point in YbRh$_2$Si$_2$*

Frank STEGLICH[1], Heike PFAU[1], Stefan LAUSBERG[1], Peijie SUN[2], Ulrike STOCKERT[1], Manuel BRANDO[1], Sven FRIEDEMANN[3], Cornelius KRELLNER[4], Christoph GEIBEL[1], Steffen WIRTH[1], Stefan KIRCHNER[1,5], Elihu ABRAHAMS[6] and Qimiao SI[7]

[1]*Max Planck Institute for Chemical Physics of Solids, 01187 Dresden, Germany*
[2]*Institute of Physics, Chinese Academy of Sciences, Beijing, China*
[3]*Cavendish Laboratory, University of Cambridge, Cambridge, UK*
[4]*Physics Institute, Goethe University, Frankfurt/Main, Germany*
[5]*Max Planck Institute for Physics of Complex Systems, 01187 Dresden, Germany*
[6]*Department of Physics and Astronomy, University of California Los Angeles, Los Angeles, California, 90095, USA*
[7]*Department of Physics and Astronomy, Rice University, Houston, Texas, 77005, USA*

E-mail: *frank.steglich@cpfs.mpg.de*

The heavy-fermion metal YbRh$_2$Si$_2$ is a weak antiferromagnet below $T_N$ = 0.07 K. Application of a low magnetic field $B_c$ = 0.06 T ($\perp$c) is sufficient to continuously suppress the antiferromagnetic (AF) order. Below $T \approx 10$ K, the Sommerfeld coefficient of the electronic specific heat $\gamma(T)$ exhibits a logarithmic divergence. At $T < 0.3$ K, $\gamma(T) \sim T^{-\varepsilon}$ ($\varepsilon$: 0.3 – 0.4), while the electrical resistivity $\rho(T) = \rho_0 + aT$ ($\rho_0$: residual resistivity). Upon extrapolating finite-$T$ data of transport and thermodynamic quantities to $T = 0$, one observes (i) a vanishing of the "Fermi surface crossover" scale $T^*(B)$, (ii) an abrupt jump of the initial Hall coefficient $R_H(B)$ and (iii) a violation of the Wiedemann Franz law at $B = B_c$, the field-induced quantum critical point (QCP). These observations are interpreted as evidence of a critical destruction of the heavy quasiparticles, i.e., propagating Kondo singlets, at the QCP of this material.

## 1. Heavy-fermion quantum criticality

Rare-earth-based intermetallic compounds with heavy-fermion (HF) phenomena ("HF metals") are well described within the framework of the Kondo lattice [1]. In contrast to other families of correlated electron materials, HF metals exhibit a clear hierarchy of relevant energy scales, i.e., spin-orbit coupling, crystal-field (CF) splitting of the localized 4*f* shell, single-ion Kondo energy and Kondo coherence [2]. Following the discovery of HF phenomena in thermodynamic and transport properties of CeAl$_3$ [3], the surprising observation of HF superconductivity in CeCu$_2$Si$_2$ [4] and subsequently in a few U-based intermetallics, like UBe$_{13}$ [5] and UPt$_3$ [6], initiated intense studies in this field. In addition to superconductivity, a second "revolution" sparked by HF research led to worldwide research activities on quantum criticality [7-9]. Antiferromagnetic (AF) quantum critical points (QCPs) in HF metals have served as prototypical settings to study non Fermi liquid (NFL) phenomena as well as quantum-criticality-driven novel phases, notably unconventional superconductivity [10].

In itinerant (*d*-electron-based) metals a spin-density-wave (SDW) QCP commonly occurs [11-14]. A SDW QCP may also be realized in HF metals, namely if the heavy (composite) charge carries behave like *d* electrons in that they keep their integrity at the QCP. For example, the HF superconductor $CeCu_2Si_2$ exhibits a three-dimensional SDW QCP [15], whose critical fluctuations were found to drive the formation of the Cooper pairs [16, 17,18].

For $CeCu_2(Si_{0.9}Ge_{0.1})_2$ two separate superconducting domes exist at different pressure ranges [19]: One of them is centered at the SDW QCP at ambient pressure (*p*), resembling the dome of superconductivity which forms around the *p*-induced QCP in $CePd_2Si_2$ [10]. The second dome occurs at about *p* = 5 GPa and is believed to be located close to a weak valence transition, where Cooper pairing is ascribed to nearly quantum critical valence fluctuations [20, 21], as expected within a "valence-crossover" QCP scenario [22, 23]. The latter has been assumed to also apply [22] to the *p*-induced AF QCP in the HF superconductor $CeRhIn_5$ [24]. For this material, de Haas-van Alphen measurements [25] performed in the field range 10 T ≤ *B* ≤ 17 T, i.e., not far above $B_{c2}$ ≈ 9 T, revealed an abrupt reconstruction of the Fermi surface (FS) to occur at $p_c$ ≈ 2.3 GPa, where the AF order is suppressed [26]. However, the fact that this suppression of the AF order occurs smoothly [26, 27] rules out the valence-crossover QCP scenario [22], as this description would require the AF quantum phase transition in $CeRhIn_5$ to be of first order. Therefore, the nature of the *p*-induced quantum criticality in $CeRhIn_5$ hints at an unconventional QCP [28-30]. In case of a "local" QCP [28, 29] strong low-dimensional spin fluctuations are assumed to cause a critical destruction of the HFs, i.e., the propagating Kondo singlets. This has led to a new low-energy scale *E\**, i.e., the finite-temperature manifestation of the abrupt FS reconstruction at the QCP (see Fig. 1a): The heavy quasiparticles exist only well below the Kondo temperature $T_K$ on the paramagnetic side where they form a heavy Landau Fermi liquid below $T_{FL}$. As here the *4f* states are delocalized, they contribute to a "large" FS. At an SDW QCP, *E\** is finite and the FS remains large (see Fig. 1b).

In the following, we discuss salient NFL phenomena as well as several pieces of evidence for a sudden FS reconstruction at the AF QCP in the tetragonal HF metal $YbRh_2Si_2$ [31-33]. This compound has been shown to behave as a prototypical Kondo lattice system [34], with a characteristic Kondo temperature $T_K$ ≈ 30 K, referring to the lowest-lying CF-derived Kramers doublet (labeled $T_0$ in Fig. 1). At $T_K$, where all $Yb^{3+}$ ions are identical, spatial coherence among the charge carriers sets in [34].

## 2. Non Fermi liquid effects in $YbRh_2Si_2$

$YbRh_2Si_2$ is a weak antiferromagnet below $T_N$ ≈ 0.07 K [31], with a tiny ordered moment $\mu_{ord}$ ≈ 2·$10^{-3}$ $\mu_B$/$Yb^{3+}$ [35]. A small magnetic field $B_c$ ≈ 0.06 T (*B*⊥c) respectively 0.66 T (*B*⊥a,b) is sufficient to suppress the AF order. This suppression occurs continuously when increasing the field through $B_c$: Measurements of the magnetostriction reveal the AF phase transition to remain of second order to the lowest accessible transition temperature of 20 mK [36]. From Fig. 2a one infers that upon cooling, the Sommerfeld coefficient of the electronic specific heat, $\gamma(T) = C_{el}(T)/T$, is diverging at the critical field $B = B_c$. In addition, below $T \approx 0.15$ K, the electrical resistivity measured at $B = B_c$ depends linearly on temperature, $\rho(T) = \rho_0 + aT$ (Fig. 2b); $\rho_0$ being the residual resistivity. Extrapolation of these $T$ dependences of $\gamma(T)$ and $\rho(T)$ to zero temperature yields a NFL ground state with diverging quasiparticle mass $m^*$ exactly at the field-induced QCP. At $B \neq B_c$, on the other hand, YbRh$_2$Si$_2$ behaves as a heavy Fermi liquid, cf. Figs. 2a and b.

In this context, several remarkable observations are worth mentioning:
- The temperature range, within which the asymptotic $\Delta\rho \sim T$ behavior is observed, extends to higher temperatures upon increasing disorder, e.g., from $T \leq 0.15$ K for the high-quality single crystal exploited in Fig. 2 ($\rho_0 \approx 0.5$ μΩcm, residual-resistivity ratio RRR $\approx 150$) to $T \leq 10$ K for YbRh$_2$(Si, Ge)$_2$, with $\rho_0 \approx 5$ μΩcm [33]. Here, $\Delta\rho = \rho(T) - \rho_0$.
- The $T$-dependence of $\rho(T)$ for 0.15 K < $T$ < 0.3 K (see Fig. 2b) and at $B = B_c$ is well described, within the framework of a new "critical Fermi liquid" theory, by $\rho = \rho'_0 + a'T^\alpha$, with $\alpha = ¾$ [37].
- Though the FS associated with the antiferromagnetically ordered phase of YbRh$_2$Si$_2$ is assumed to be *small* (as argued below), the low-field Fermi liquid phase appears to be particularly *heavy*: Its quasiparticle mass even exceeds the one in the paramagnetic Fermi liquid phase (see Fig. 2a). While surprising at first sight, this observation is attributed to the *dynamical* Kondo screening [38], which determines thermodynamics – although the *static* Kondo effect is absent in the ground state.

## 3. Discontinuous reconstruction of the Fermi surface at the AF QCP in YbRh$_2$Si$_2$

Direct FS studies at the QCP in YbRh$_2$Si$_2$ are not possible. Angle-resolved photoemission spectroscopy (ARPES) below $T = 0.1$ K with a correspondingly high energy resolution is not available at present. On the other hand, measurements of magnetic quantum oscillations require the application of magnetic fields of several T, which would fully suppress the quantum critical fluctuations in YbRh$_2$Si$_2$ [9]. Therefore, in order to investigate the evolution of the FS in the quantum critical regime of this material one has to rely on thermodynamic and transport measurements.

Isothermal studies of the magnetic field dependence of the initial Hall coefficient $R_H$ (being almost identical with the *normal* Hall coefficient at $T < 1$ K) revealed a large drop in $R_H(B)$ at a crossover field $B^*(T)$ and a substantial narrowing of this crossover upon cooling [40]. Fig. 3a displays the crossover line $T^*(B)$ [$=T(B^*)$], which is found to merge with both $T_N(B)$ and $T_{FL}(B)$ at $B = B_c$ in the zero-temperature limit [41] and has been shown [36] to represent a new thermodynamic energy scale [28, 29]. Careful investigations of single crystals of widely differing quality showed, via various types of magnetotransport probes, that the crossover width is proportional to temperature [41] (cf. Fig. 3b). This implies, upon extrapolation to $T = 0$, an abrupt finite jump of $R_H(B)$ at the field-induced QCP and verifies the nature of the Kondo breakdown QCP as predicted by Si et al. [28] and Coleman et al. [29]. In the following, we shall address the dynamical processes associated with this unique type of instability.

## 4. Violation of the Wiedemann Franz law at the AF QCP in YbRh$_2$Si$_2$

The Wiedemann Franz (WF) law describes the combined heat and charge transport in a metal at absolute zero temperature, where all scatterings are elastic. Defining the thermal resistivity by $w = L_0 T/\kappa$, where $\kappa$ is the thermal conductivity and $L_0 = (\pi k_B)^2/3e^2$ Sommerfeld's constant, the WF law states that (as $T \to 0$) the residual thermal and electrical resistivities are identical: $\rho_0/w_0 = L/L_0 = 1$. Here, $L = \rho\kappa/T$ denotes the Lorenz number, while $L/L_0$ is called the Lorenz ratio. The WF law, one of the fundamental laws in metal physics, can in principle be violated in two very different ways: If, in addition to the electronic quasiparticles, charge-neutral fermionic excitations ("spinons") contribute to the heat current but not to the charge current, the Lorenz ratio $L/L_0 > 1$, even in the zero-temperature limit. On the other hand, a breakdown of Landau's quasiparticle concept may lead to $L/L_0 < 1$ at $T = 0$.

In Fig. 4 the low-$T$ behavior of $\rho(T)$ and $w(T)$ is displayed for a single-crystalline YbRh$_2$Si$_2$ sample of medium quality (RRR $\approx$ 40) at several different values of the control parameter, magnetic field $B$ [42]. For the highest applied fields of 0.6 T (Fig. 4i) and 1 T (Fig. 4j), both $\rho(T)$ and $w(T)$ show a very strong $T^2$ dependence at sufficiently low temperature, characteristic of a heavy Fermi liquid phase. Extrapolating these results to $T = 0$, one recognizes $w_0 = \rho_0$ within the experimental uncertainty; i.o.w., the WF law holds even in case of such an extremely heavy Landau Fermi liquid phase.

In contrast, at $B = 0$ both the electrical and thermal resistivities exhibit a linear temperature dependence above, respectively, $T_N = 0.07$ K and $T \approx 0.12$ K, where $w(T)$ starts to drop (see Fig. 4a). Below the Néel temperature, one finds $\rho = \rho_0 + AT^2$ with a huge value of the coefficient $A$. At the lowest accessible temperature of $\approx 25$ mK the thermal resistivity is clearly smaller than its electrical counterpart: $w < \rho$. This proves that, in addition to the electronic ones, another species of heat carriers is present at finite temperatures. These are the acoustic AF magnons, as identified with the aid of low-$T$ specific heat results, see Fig. 5 [33]: The magnon specific heat obeys $C_m \sim T^3$ below $T \approx 0.05$ K, which implies a magnon thermal conductivity $\kappa \sim T^3$ at sufficiently low temperatures, too. At $T = 0$, the heat current is carried exclusively by the electronic quasiparticles. Because of the Fermi liquid phase, the WF law must hold in the antiferromagnetically ordered state ($w_0 = \rho_0$). Unfortunately, this cannot be observed directly as the electronic heat transport is fully masked by the bosonic one, due to magnons below $T_N = 0.07$ K and due to short-lived magnon excitations ("paramagnons") in the temperature window $T_N < T < 0.12$ K at $B = 0$. At finite $B < B_c$ an analogous behavior is observed (Fig. 4b, $B = 0.02$ T). This paramagnon contribution is also visible below $T \approx 0.07$ K at $B \approx B_c$ (Fig. 4c), but becomes suppressed at sufficiently high magnetic fields, $B \geq 0.2$ T (Figs. 4f-j).

The values of the residual electrical and electronic thermal resistivities at $B = B_c$, $\rho_0$ and $w_0$, are obtained by extrapolating the NFL-type linear-in-$T$ dependences of $\rho(T)$ and $w(T)$ to zero temperature, cf. Supplementary Information of Ref. 42. Such an extrapolation is necessary because of the influence of the above-mentioned paramagnons. It yields $w_0 > \rho_0$ or $L/L_0 < 1$, i.e., a violation of the WF law, exactly at the QCP.

Very similar experimental data have recently been reported by two groups [43, 44], who studied YbRh$_2$Si$_2$ single crystals of considerably higher quality (RRR ≳ 100) compared to our sample. As is evident from Refs. 43 and 44, the paramagnon heat transport in cleaner samples appears to persist at higher fields. This, along with their limitation in applying large enough magnetic fields (compared to $B_c$ of the respective sample orientation), prevents the authors of Refs. 43 and 44 to observe both the validity of the WF law in the paramagnetic phase and the field-induced suppression of the bosonic contribution to the heat transport. Therefore, they attribute the downturn in $w(T)$ from the linear-in-$T$ behavior seen at $B = B_c$ to the *electronic* heat carriers and claim the WF law to hold at the field-induced QCP in YbRh$_2$Si$_2$.

In order to support our conclusion that the WF law is indeed violated as $T \to 0$ at $B = B_c$, we show in Fig. 6 the isothermal field dependence $L(B)/L_0$ for 0.1 K ≤ $T$ ≤ 0.4 K. Data at lower temperatures have been ignored because of the interfering bosonic contribution in the low-field range as discussed above. This means that all $L(B)/L_0$ data in Fig. 6 are representative of purely electronic transport. Except for the results taken at $T = 0.1$ K and the highest fields of 0.6 T and 1 T which, within the error bars, are close to the "WF value" $L/L_0 = 1$, all Lorenz ratios are smaller than 1. This indicates that the electronic heat carriers are subject to dominating inelastic scatterings, i.e., from both AF spin fluctuations and electronic charge carriers. At zero temperature these scatterings are frozen out, which is in accordance with the validity of the WF law in a Fermi liquid with sharp FS, e.g., at both $B < B_c$ and $B > B_c$ for YbRh$_2$Si$_2$.

In the quantum critical regime, where the FS is fluctuating, the quasiparticle weights taken at the respective small and large values of the Fermi wave vector satisfy dynamical, $\omega/T$, scaling and smoothly vanish exactly at the QCP [42], cf. Fig. 7. This warrants the critical FS fluctuations to exist in the whole temperature range of quantum critical behavior, $0 \leq T \lesssim 1$ K. They represent a novel type of quantum critical fluctuations, i.e., they are fermionic in origin and operate as additional inelastic scatterers for the electronic quasiparticles. These FS fluctuations give rise to the distinct minimum in the $L(B)/L_0$ isotherms close to $T^*(B)$ (Fig. 6), which highlights an intimate relationship between those additional inelastic scatterings and the Fermi surface crossover.

Recent results of isothermal measurements of $\kappa(B)$ and $\rho(B)$, performed on a high-quality YbRh$_2$Si$_2$ single crystal (RRR ≈ 150) at $T = 0.49$ K [45] fit well to those obtained from the $T$-scans on our medium-quality sample with RRR ≈ 40, cf. Fig. 6. This proves that the minimum in the $L(B)/L_0$ isotherms is a robust feature which, like the Hall crossover, becomes considerably narrower upon cooling (Fig. 6). As a consequence of the $\omega/T$ scaling, the additional electronic inelastic scatterings extend to $\omega = 0$ exactly at the QCP. This provides a very natural explanation for our main result that $L(T \to 0)/L_0 < 1$ at $B = B_c$ and $L(T \to 0)/L_0 = 1$ at $B \neq B_c$. The fact that we find a reduction of the Lorenz ratio by "only" 10% is in full accord with the generalized quasiparticle-quasiparticle nature of the underlying scatterings. Though many-body in origin they include a finite, moderate fraction of small-angle-scattering processes - in analogy to the electron-electron scatterings in simple metals [46].

The data presented in Fig. 8 lend further support to our conclusion of the WF law being violated at the QCP in YbRh$_2$Si$_2$: The electrical resistivity $\rho(T)$, measured on the medium-quality single crystal used for our study of the heat conductivity, clearly displays that the sample is heated up only in the vicinity of $B_c$ = 0.059 T, namely, if an appropriately large current is applied at low temperatures [47]. Obviously, an extra Joule's heat is generated here which, despite of the additional bosonic heat carriers, cannot be properly lead away. Very likely, this extra heat results from the presence of particularly strong inelastic scatterings, i.e., those discussed before. Some heating effect is still visible at $B$ = 0.06 T even when a rather low current is injected into the sample (see black trace). Upon proper extrapolation of these low-current data from the regime where $\Delta\rho \sim T$ to $T$ = 0, the residual resistivity at $B$ = 0.06 T is found to be almost identical to the $\rho_0$ values extrapolated for both $B$ = 0.1 T and 0.2 T. As a result, $\rho_0(B)$ is found to jump abruptly from a larger to smaller value upon raising the field through $B_c$, manifesting an abrupt increase in the charge carrier concentration as already inferred from the Hall measurements discussed in Section 3 [40, 41, 48].

Recently, $L(B)/L_0$ was also determined for the HF metal YbAgGe [49]. When the results taken at $B$ = 4.5 T are extrapolated to $T$ = 0, where a nearby bi-quantum critical point is anticipated [50], $L_{el}/L_0 \approx 0.92$ is obtained. YbRh$_2$Si$_2$ and YbAgGe, therefore, appear to be the first, and so far only, metals for which this type of fundamental violation of the WF law could be established [51].

## 5. Perspective

In this paper, we have presented and discussed three pieces of evidence for a field-induced Kondo destroying AF QCP in YbRh$_2$Si$_2$. Upon reliably extrapolating our data taken at finite temperature to $T$ = 0, we recognize exactly at $B = B_c$:
(i) a merging of the quantum-critical energy scale $T^*(B)$ with the magnetic phase boundary $T_N(B)$ and the crossover line $T_{FL}(B)$, (ii) an abrupt jump of the initial Hall coefficient $R_H(B)$ and (iii) a violation of the WF law. While the first result locates the Kondo breakdown in the phase diagram and the second one quantifies the strength of this phenomenon, the latter result illustrates the dynamical processes, causing the apparent breakup of the Kondo singlets. Future research will focus on studies of this Mott-type instability in the absence of any interfering magnetism as expected for, e.g., Yb(Rh$_{1-x}$Ir$_x$)$_2$Si$_2$ with $x \gtrsim 0.1$ [56].

Another promising area of future investigations concerns the occurrence of a *ferromagnetic* QCP in HF metals, like YbNi$_4$(P$_{1-x}$As$_x$)$_2$, $x \approx 0.1$ [57]. This type of instability does apparently not exist in itinerant (*d*-electron-based) metallic materials [58]. On the other hand, it was shown [59] that for Kondo lattice systems the Kondo effect can be fully suppressed inside a ferromagnetically ordered phase. It has yet to be explored, whether in YbNi$_4$(P$_{0.9}$As$_{0.1}$)$_2$ the breakdown of the Kondo effect coincides with the ferromagnetic QCP, in analogy to the observation in the weak antiferromagnet YbRh$_2$Si$_2$.

**Acknowledgments**

Valuable conversations with E. Bauer, S. Bühler-Paschen, P. Coleman, R. Daou, H. Fukuyama, S. Hartmann, K. Kanoda, J.-Ph. Reid, J. Schmalian, L. Taillefer, H. von Löhneysen, P. Wölfle and G. Zwicknagl are gratefully acknowledged. The work performed at the MPI for Chemical Physics of Solids was partly supported by the DFG under the auspices of FOR 960 "Quantum Phase Transitions". The work at Rice University was in part supported by the NSF Grant No. DMR – 1309531 and the Robert A. Welch Grant No. C – 1411.

**References**

[1] S. Doniach: Physics B+C **91** (1977) 231.
[2] N. Grewe and F. Steglich: in *Handbook on the Physics and Chemistry of Rare Earths*, ed. K. A. Gschneidner, J. L. Eyring (Elsevier, Amsterdam, 1991) Vol. 14, p. 343.
[3] K. Andres, J. E. Graebner and H.R. Ott: Phys. Rev. Lett. **35** (1975) 1779.
[4] F. Steglich, J. Aarts, C. D. Bredl, W. Lieke, D. Meschede, W. Franz and H. Schäfer: Phys. Rev. Lett. **43** (1979) 1892.
[5] H. R. Ott, H. Rudiger, Z. Fisk and J. L. Smith: Phys. Rev. Lett. **50** (1983) 1595.
[6] G. R. Stewart, Z. Fisk, J. O. Willis and J. L. Smith: Phys. Rev. Lett. **52** (1984) 679.
[7] G. R. Stewart: Rev. Mod. Phys. **73** (2001) 797; **78** (2006) 743.
[8] H. von Löhneysen, A. Rosch. M. Vojta and P. Wölfle: Rev. Mod Phys. **79** (2007) 1015.
[9] P. Gegenwart, Q. Si and F. Steglich: Nature Phys. **4** (2008) 186.
[10] N. D. Mathur, F. M. Grosche, S. R. Julian, I. R. Walker, D. M. Freye, R. K. W. Haselwimmer and G. G. Lonzarich: Nature **394** (1998) 39.
[11] J. A. Hertz: Phys. Rev. B **14** (1976) 1165.
[12] T. Moriya: *Spin Fluctuations in Itinerant Electron Magnetism* (Springer, Berlin 1985).
[13] A. J. Millis: Phys. Rev. B **48** (1993) 7183.
[14] M. A. Continentino, G. M. Japiassu and A. Troper: Phys. Rev. B **39** (1989) 9734.
[15] J. Arndt, O. Stockert, H. S. Jeevan, C. Geibel, F. Steglich, E. Faulhaber, M. Loewenhaupt, K. Schmalzl and W. Schmidt: Phys. Rev. Lett. **106** (2011) 246401.
[16] O. Stockert, J. Arndt, E. Faulhaber, C. Geibel, H. S. Jeevan, S. Kirchner, M. Loewenhaupt, K. Schmalzl, W. Schmidt, Q. Si and F. Steglich: Nature Phys. **7** (2011) 119.
[17] O.Stockert, S. Kirchner, F. Steglich and Q. Si: J. Phys. Soc. Jpn. **81** (2012) 011001.
[18] The exchange-energy saving, extracted from the difference between the dynamical susceptibilities of the normal and superconducting states, largely exceeds the superconducting condensation energy (by a factor of about 20). This has been interpreted in terms of quantum critical fluctuations beyond the SDW description at intermediate energies and temperatures [16, 17].
[19] H. Q. Yuan, F. M. Grosche, M. Deppe, C. Geibel, G. Sparn and F. Steglich: Science **302** (2003) 2104.
[20] K. Miyake, O. Narikiyo and Y. Onishi: Physica B **259-261** (1999) 676; Y. Onishi and K. Miyake: J. Phys. Soc. Jpn. **69** (2001) 3955; A. Holmes, D. Jaccard and K. Miyake: Phys. Rev. B **69** (2004) 024508.
[21] P. Monthoux and G. G. Lonzarich: Phys. Rev. B **69** (2004) 064517.
[22] S. Watanabe and K. Miyake: J. Phys. Soc. Jpn. **82** (2013) 083704.
[23] The interpretation given for CeCu$_2$Si$_2$ in [20, 21] is questioned by: L. V. Pourosvskii, P. Hansmann, M. Ferrero and A. Georges, arXiv: 1305.5204 vl [cond mat. str-el] 22 May 2013.
[24] H. Hegger, C. Petrovic, E. G. Moshopoulou, M. F. Hundley, J. L. Sarrao, Z. Fisk and J. D. Thompson:

Phys. Rev. Lett. **84** (2000) 4986.
[25] H. Shishido, R. Settai, H. Harima and Y. Ōnuki: J. Phys. Soc. Jpn. **74** (2005) 1103.
[26] T. Park, F. Ronning, H. Q. Yuan, M. B. Salamon, R. Movshovich, J. L. Sarrao and J. D. Thompson: Nature **440** (2006) 65.
[27] G. Knebel, D. Aoki, D. Braithwaite, B. Salce and J. Flouquet: Phys. Rev. B. **74** (2006) 020501(R).
[28] Q. Si, S. Rabello, K. Ingersent and J. L. Smith: Nature **413** (2001) 804.
[29] P. Coleman, C. Pépin. Q. Si and R. Ramazashvili: J. Phys.: Condens. Matter **13** (2001) R723.
[30] T. Senthil, M. Vojta and S. Sachdev: Phys. Rev. B. **69** (2004) 035111.
[31] O. Trovarelli, C. Geibel, S. Mederle, C. Langhammer, F. M. Grosche, P. Gegenwart, M. Lang, G. Sparn and F. Steglich: Phys. Rev. Lett. **85** (2000) 626.
[32] P. Gegenwart, J. Custers, C. Geibel, K. Neumaier, T. Tayama, K. Tenya, O. Trovarelli and F. Steglich: Phys. Rev. Lett. **89** (2002) 056402.
[33] J. Custers, P. Gegenwart, H. Wilhelm, K. Neumaier, Y. Tokiwa, O. Trovarelli, C. Geibel, F. Steglich, C. Pépin and P. Coleman: Nature **424** (2003) 524.
[34] S. Ernst, S. Kirchner, C. Krellner, C. Geibel, G. Zwicknagl, F. Steglich and S. Wirth: Nature **474** (2011) 362.
[35] K. Ishida, D. E. MacLaughlin, Ben-Li Young, K. Okamoto, Y. Kawasaki, Y. Kitaoka, G. J. Niewenhuys. R. H. Heffner, O. O. Bernal, W. Higemoto, A. Koda, R. Kadono, O. Trovarelli, C. Geibel and F. Steglich: Phys. Rev. B. **68** (2003) 184401.
[36] P. Gegenwart, T. Westerkamp, C. Krellner, Y. Tokiwa, S. Paschen, C. Geibel, F. Steglich, E. Abrahams and Q. Si: Science **315** (2007) 969.
[37] P. Wölfle and E. Abrahams: Phys. Rev. B. **84** (2011) 041101(R).
[38] J.-X. Zhu, D. Grempel and Q. Si: Phys. Rev. Lett. **91** (2003) 156404. For a recent review on this issue, see Section 3.2 of Ref. 39.
[39] Q. Si and S. Paschen: Phys. Status Solidi B **250** (2013) 425.
[40] S. Paschen, T. Lühmann, S. Wirth, P. Gegenwart, O. Trovarelli, C. Geibel, F. Steglich, P. Coleman and Q. Si: Nature **432** (2004) 881.
[41] S. Friedemann, N. Oeschler, S. Wirth, C. Krellner, C. Geibel, F. Steglich, S. Paschen, S. Kirchner and Q. Si: Proc. Natl. Acad. Sci. USA **107** (2010) 14547.
[42] H. Pfau, S. Hartmann, U. Stockert. P. Sun, S. Lausberg, M. Brando, S. Friedemann, C. Krellner, C. Geibel, S. Wirth, S. Kirchner, E. Abrahams, Q. Si and F. Steglich: Nature **484** (2012) 493.
[43] Y. Machida, K. Tomokuni, K. Izawa, G. Lapertot, G. Knebel. J.-P. Brison and J. Flouquet: Phys. Rev. Lett. **110** (2013) 236402.
[44] J.-Ph. Reid, M.A. Tanatar, R. Daou, Rongwei Hu, C. Petrovic and L. Taillefer: arXiv:1309.6315v1 [cond-mat.str-el] 24 Sep 2013.
[45] H. Pfau, R. Daou, S. Lausberg, H. R. Naren, M. Brando, S. Friedemann, S. Wirth, T. Westerkamp, U. Stockert, P. Gegenwart, C. Krellner, C. Geibel, G. Zwicknagl and F. Steglich: Phys. Rev. Lett. **110** (2013) 256403.
[46] J. M. Ziman: *Electrons and Phonons* (Oxford University Press, Oxford, 1960).
[47] S. Lausberg, Dissertation, TU Dresden, 2013 (unpublished).
[48] S. Friedemann, S. Wirth, N. Oeschler, C. Krellner, C. Geibel, F. Steglich, S. MaQuilon, Z. Fisk, S. Paschen and G. Zwicknagl: Phys. Rev. B **82** (2010) 035103.
[49] J. K. Dong, Y. Tokiwa, S. L. Bud'ko, P. C. Canfield and P. Gegenwart: Phys. Rev. Lett. **110** (2013) 176402.
[50] Y. Tokiwa, M. Garst, P. Gegenwart, S. L. Bud'ko and P. C. Canfield: arXiv: 1306.4251v2 [cond-mat.str-el] 21 June 2013.
[51] For the tetragonal, quasi-two-dimensional HF metal CeCoIn$_5$, for which an AF QCP is suspected [52] but not identified, the Lorenz ratio $L/L_0$ was extrapolated to about 0.8 as $T \rightarrow 0$ for c-axis transport, while it approaches $L/L_0 = 1$ for in-plane transport [53]. This finding was discussed in terms of putative strongly anisotropic quantum critical fluctuations although, in the zero-temperature limit, spin fluctuations are unapt to raise $w_0$ relative to $\rho_0$ [54]. Subsequently, these results were consistently

explained within the framework of quasi-two-dimensional transport [55].

[52] S. Zaum, K. Grube, R. Schäfer, E. D. Bauer, J. D. Thompson and H. von Löhneysen: Phys. Rev. Lett. **106** (2011) 087003.
[53] M. A. Tanatar, J. Paglione, C. Petrovic and L. Taillefer: Science **316** (2007) 1320.
[54] See, e.g., R. P. Smith, M. Sutherhand, G. G. Lonzarich, S. S. Saxena, N. Kimura, S. Takashima, M. Nohara and H. Takagi: Nature **455** (2008) 1220.
[55] M. F. Smith and R. H. McKenzie: Phys. Rev. Lett. **101** (2008) 266403.
[56] S. Friedemann, T. Westerkamp, M. Brando, N. Oeschler, S. Wirth, P. Gegenwart, C. Krellner, C. Geibel and F. Steglich: Nature Phys. **5** (2009) 465.
[57] A. Steppke, R. Küchler, S. Lausberg, E. Lengyel, L. Steinke, R. Borth, T. Lühmann, C. Krellner, M. Nicklas, C. Geibel, F. Steglich and M. Brando: Science **339** (2013) 933.
[58] D. Belitz, T. Kirkpatrick and T. Vojta: Phys. Rev. Lett. **82** (1999) 4707.
[59] S. J. Yamamoto and Q. Si: Proc. Natl. Acad. Sci. USA **107** (2010) 15704.

**Figure Captions**

Fig. 1. Schematic view of the temperature ($T$) - control parameter ($\delta$) phase diagram near antiferromagnetic (AF) quantum critical points (QCPs) of Kondo break-down (a) and spin-density-wave (b) types. $T_0$ indicates the onset of local Kondo screening, once all Kondo ions have adapted the lowest lying, crystal-field derived Kramers doublet. $E^*$ is a quantum critical energy scale indicating a Fermi surface crossover from "small" to "large". From Ref. 9.

Fig. 2. Low-temperature thermodynamic and transport properties near the QCP in YbRh$_2$Si$_2$. (a) Sommerfeld coefficient of the electronic specific heat $\gamma$ vs $T$ at three magnetic fields applied within the basal, tetragonal plane ($\perp$c). At the critical field ($B_c \approx 0.06$ T), one observes $\gamma \sim T^{-\varepsilon}$ ($\varepsilon \approx 0.3 - 0.4$) at the lowest temperatures and $\gamma \sim \ln(T_0/T)$ at elevated temperatures ($T_0 \approx 24$ K, i.e., close to $T_K \approx 30$ K). Inset: $\gamma$ vs $T$ at zero field over an extended $T$ range. (b) Electrical resistivity $\rho$ vs $T$ at the same fields as in (a). From Ref. 9.

Fig. 3. (a) Position of the Fermi surface crossover in various magneto-transport experiments on samples of different quality in the temperature-field phase diagram of

YbRh$_2$Si$_2$. Red horizontal bars are crossover widths, cf. (b). Dotted/dashed line marks the magnetic phase boundary $T_N(B)$/crossover to paramagnetic Landau-Fermi liquid phase $T_{FL}(B)$. (b) Crossover width (full width at half maximum, FWHM) obtained from the same measurements exploited in (a). From Ref. 41.

Fig. 4. Thermal resistivity $w(T) = L_0 T/\kappa(T)$ (red) and electrical resistivity $\rho(T)$ (blue) below $T = 0.5$ K for $B = 0$ (a), 0.02 T (b), 0.06 T (c), 0.08 T (d), 0.1 T (e), 0.2 T (f), 0.3 T (g), 0.4 T (h), 0.6 T (i) and 1 T (j), $B \perp c$. Arrows indicate crossover to Fermi liquid ($\rho - \rho_0 = AT^2$) behavior. Representative error bars are shown for a few selected temperatures. Dashed lines in (a) – (c) indicate linear regimes used for extrapolation to $T = 0$. From Ref. 42.

Fig. 5. Specific heat of YbRh$_2$Si$_2$ as $\Delta C/T$ vs $T^2$. $\Delta C(T) = C(T) - C_{ph}(T) - C_Q(T)$, where $C_{ph}$ ($C_Q$) denotes the phonon (nuclear-quadrupole) contribution. Red line indicates a $T^3$ contribution to $\Delta C(T)$ below $T \approx 0.05$ K. From Ref. 33.

Fig. 6. Isothermal field scans of the Lorenz ratio $L(B)/L_0$ at varying temperatures 0.1 K ≤ $T$ < 0.5 K. From Refs. 42 and Ref. 45. Inset: Data from isothermal field scans of the thermal conductivity and electrical resistivity at $T = 0.49$ K and $B \leq 12$ T. From Ref. 45.

Fig. 7. Collapse of the quasiparticle weights across the local QCP. $Z_L$ and $Z_S$ are the quasiparticle weights for the "small" (left inset) and "large" (right inset) Fermi surfaces, respectively. At the QCP, the quasiparticles are critical on both the small and the large Fermi surfaces. From Ref. 42.

Fig. 8. Temperature dependence of the electrical resistivity below $T = 0.1$ K with two different excitation currents (black and red traces, respectively) at several magnetic fields. Where black and red traces overlap, the red ones are extrapolated (in green) to $T = 0$ assuming a $\Delta\rho \sim T^2$ dependence. The splitting between red and black traces close to the critical field, $B_c = 0.059$ T, below $T = 0.07$ K (marked by arrows) illustrates a heating of the sample under the applied larger current. This illustrates the violation of the Wiedemann Franz law at the QCP in YbRh$_2$Si$_2$. A sample heating below $T = 0.06$ K is still visible at $B = 0.06$ T under the smaller current, cf. extrapolated dashed straight

line. Horizontal blue hatching displays an almost field-independent residual ($T \to 0$) resistivity in the paramagnetic regime. The weak field dependence of $\rho_0$ inside the antiferromagnetically phase [41] is masked by the width of the grey hatching. The difference between the hatched horizontal regions indicates an abrupt decrease in $\rho_0(B)$ upon increasing the field through $B_c$. This illustrates a corresponding abrupt increase in the charge-carrier concentration at $T = 0$. From Ref. 47.

Fig 1

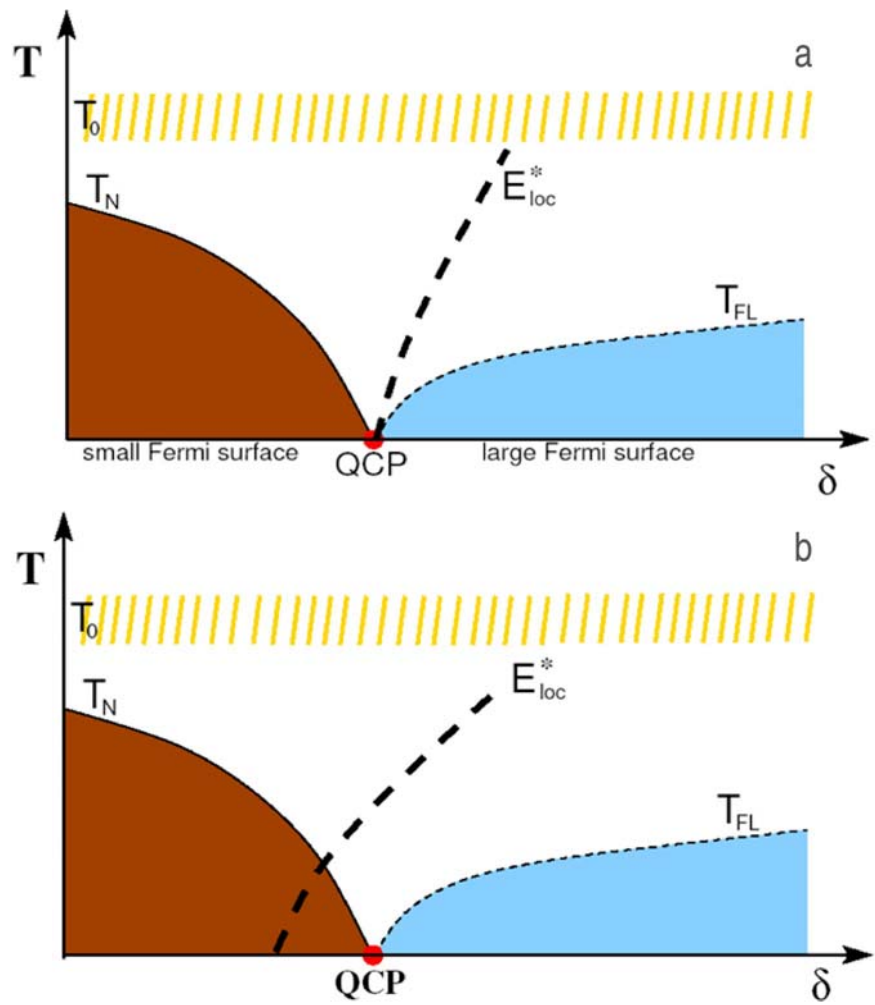

Fig. 2

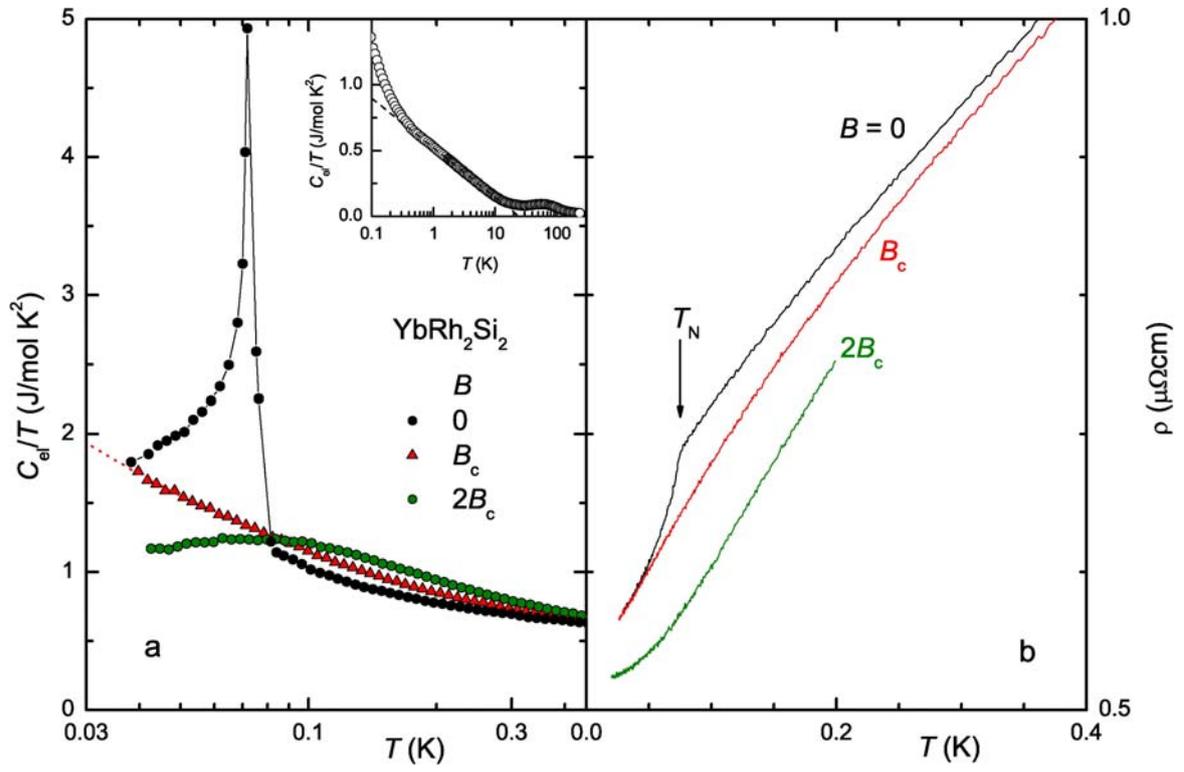

Fig. 3

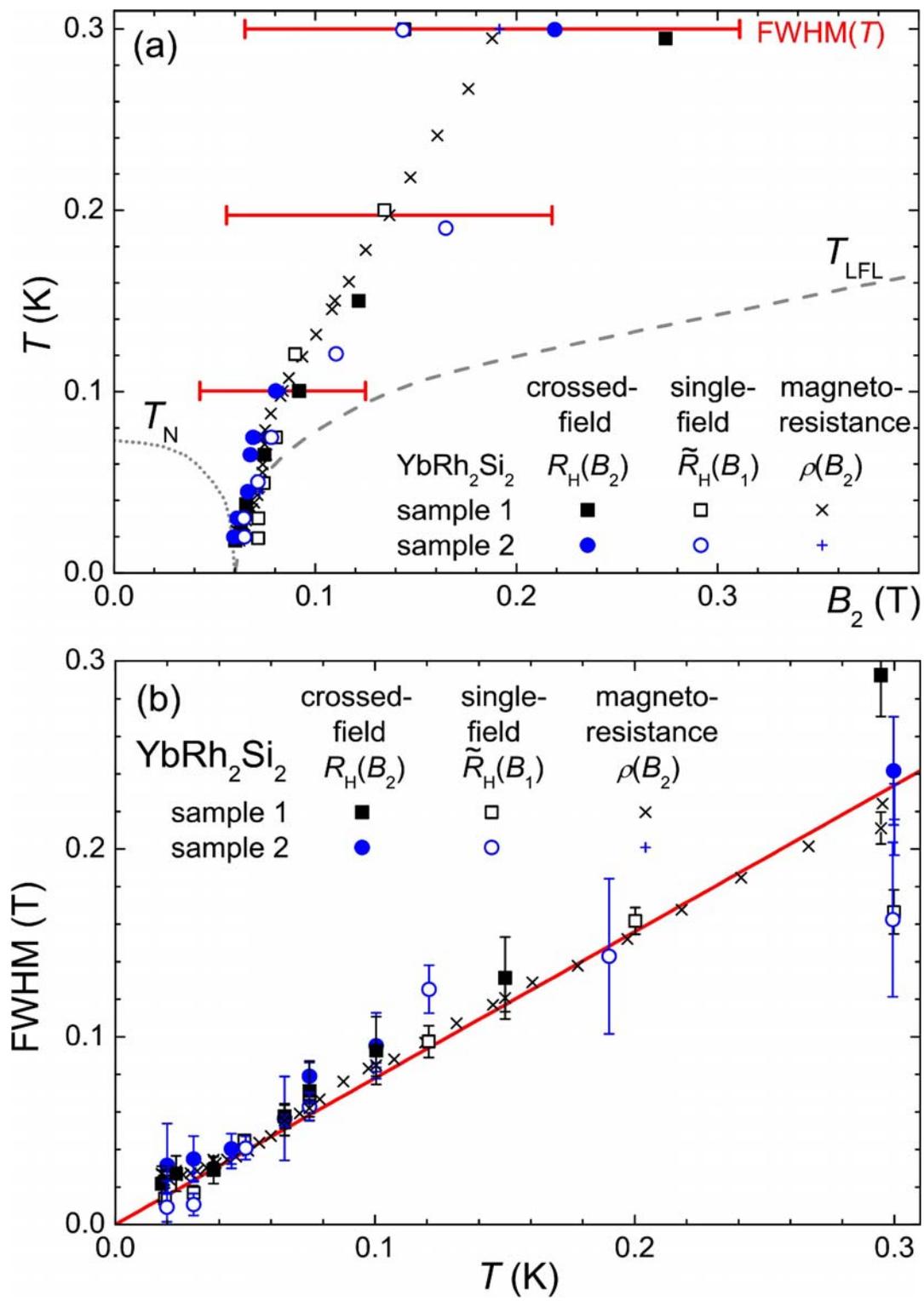

Fig. 4

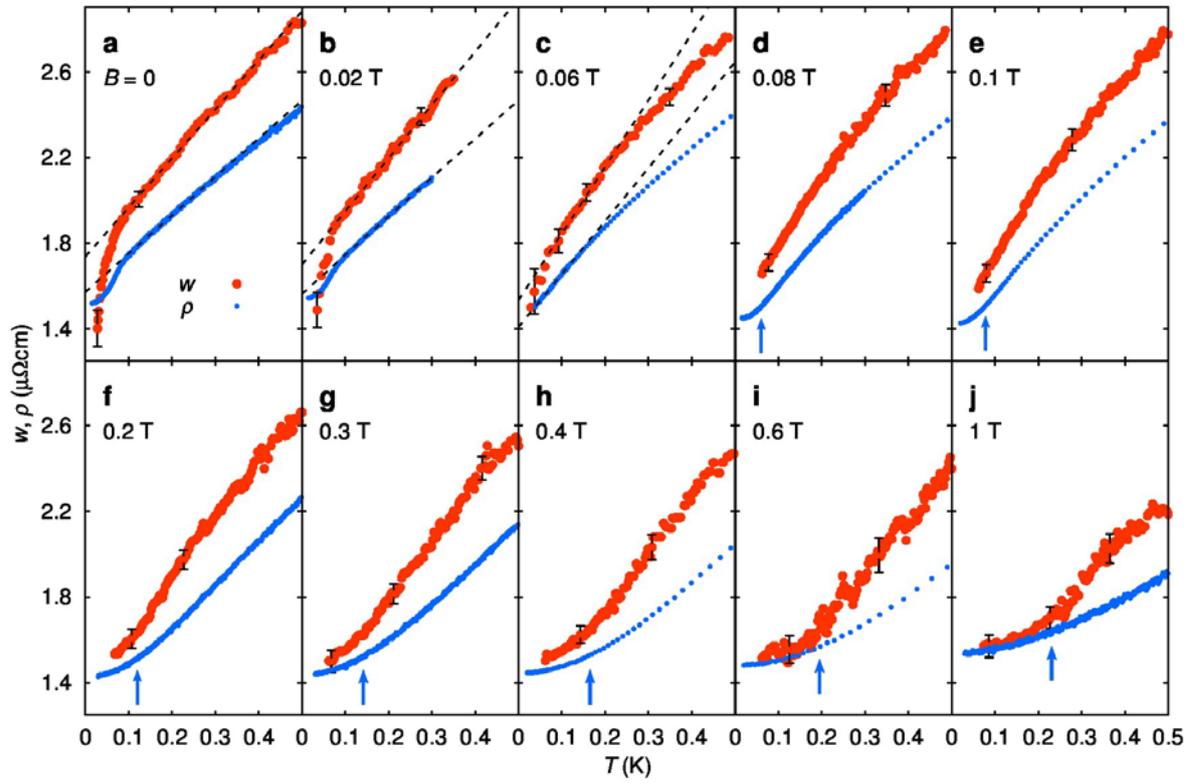

Fig. 5

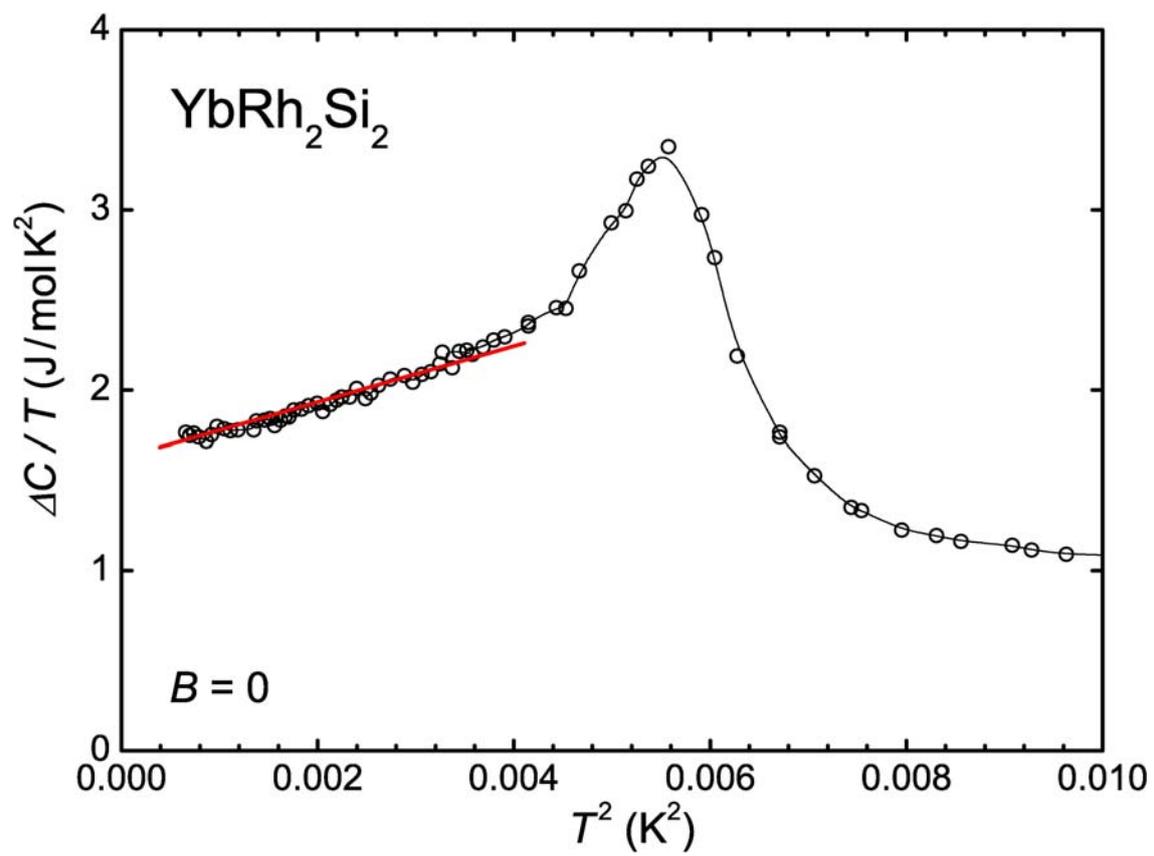

Fig. 6

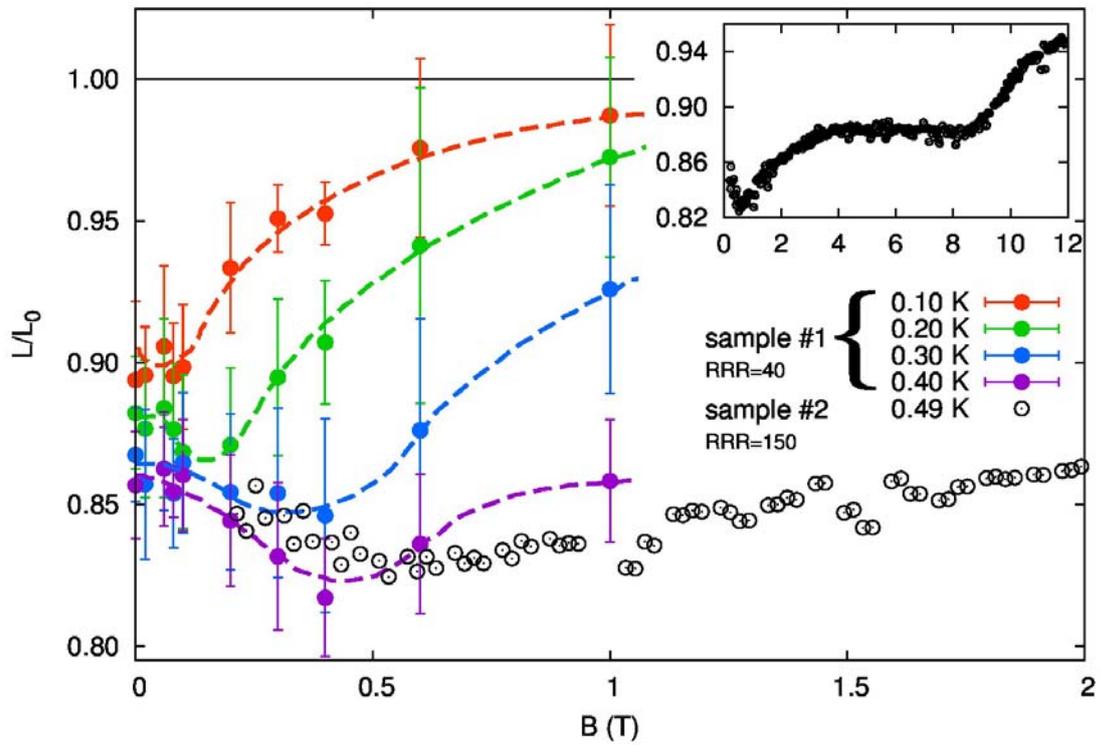

Fig. 7

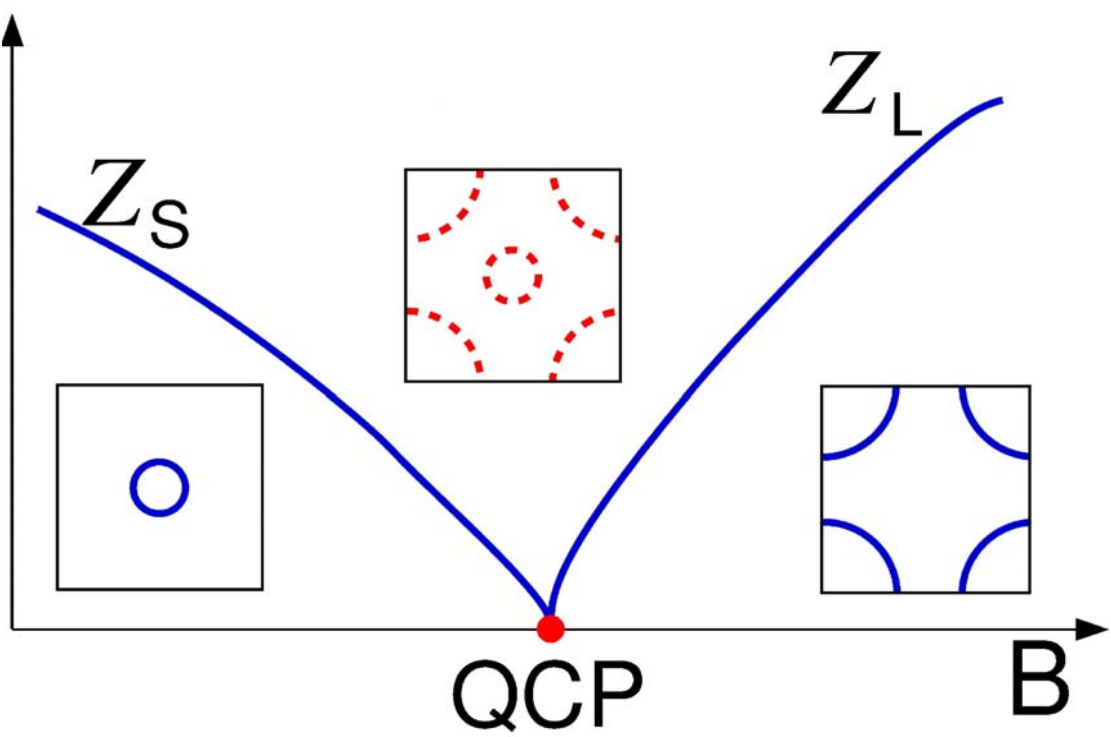

Fig. 8

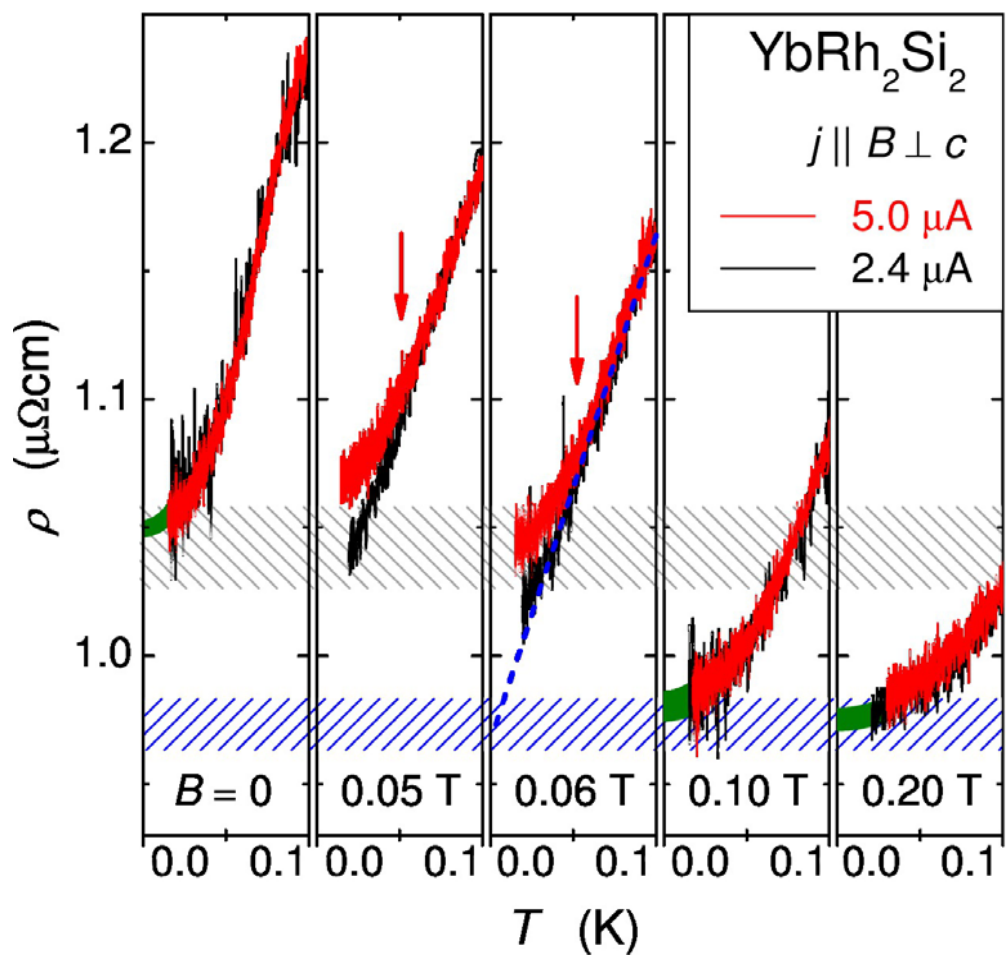